\documentclass[twocolumn,showpacs,preprintnumbers,amsmath,amssymb,aps]{revtex4}

\usepackage{graphicx}
\usepackage{dcolumn}
\usepackage{bm}

\begin{document}

\preprint{}

\title{Oscillatons formed by non linear gravity}

\author{Octavio Obreg\'on}
 \email{octavio@fisica.ugto.mx}
\author{L. Arturo Ure\~na--L\'opez}%
 \email{lurena@fisica.ugto.mx}
\affiliation{%
Instituto de F\'isica de la Universidad de Guanajuato, A.P. E-143,
C.P. 37150, Le\'on, Guanajuato, M\'exico.}%

\author{Franz E. Schunck}
 \email{fs@thp.uni-koeln.de}
\affiliation{
Institut f\"ur Theoretische Physik, Universit\"at zu K\"oln,
50923 K\"oln, Germany.
}%

\date{\today}

\begin{abstract}
Oscillatons are solutions of the coupled Einstein-Klein-Gordon (EKG)
equations that are globally regular and asymptotically flat.
By means of a Legendre transformation we are able to visualize the
behavior of the corresponding objects in non-linear gravity where
the scalar field has been absorbed by means of the conformal mapping.
\end{abstract}

\maketitle

\section{Introduction}
Non-linear modifications of the Einstein-Hilbert action have a long
history\cite{lanczos}(for a discussion on recent issues
see\cite{flanagana,flanaganb} and references therein).  They have been
of interest for a variety of reasons. For instance, it has been
claimed, that they could be good renormalizable models for quantum
gravity \cite{pimentela,pimentelb,pimentelc}. Also some non-linear
Lagrangians can be chosen with the property that the field equations
of the metric are second order, these are the so-called Lovelock
actions\cite{pimentel2a,pimentel2b,pimentel2c,pimentel2d,pimentel2e},
which arise from dimensional reduction of the Euler
characteristic. Also string theory predicts an effective gravitational
action containing the usual Einstein term plus higher-order
perturbative corrections in the
curvature\cite{pimenpola,pimenpolb}. When the Lagrangian is an exact
function of the Ricci scalar, a mapping exists to the usual
Hilbert-Einstein Lagrangian with a specific self-interacting real
scalar field; the equations of motion of the latter being the so
called Einstein-Klein-Gordon (EKG) system. This idea has been used, in
different manners to relate well-known scalar inflationary potentials
with pure-gravity higher-order curvature scalar
Lagrangians\cite{potentialsa,potentialsb,potentialsc,potentialsf,potentialsg,potentialsh}.

On the other hand, the EKG system has been studied in many situations. In particular, it has been shown that there are spherically symmetric solutions that are \emph{globally regular} and asymptotically flat; these solutions are called \emph{oscillatons}\cite{seidel91,seidel94,laul2002a,laul2002b,phi2003,laul2003a,don2003}\footnote{If instead a {\em complex} scalar is the Klein-Gordon field, boson star, boson halo or non-topological soliton solutions occur which are stable, see the most recent review on boson stars in\cite{SM03}}.

Our aim in this paper is to investigate the mapping  of
oscillatons to non-linear gravity (NLG) theories. We shall search
for the corresponding objects in these theories, and whether we
can get interesting information  relating the results in the NLG
theory and those in the EKG system.

We will limit our models to specific scalar potentials for which
NLG theories can be constructed, and we will study the "mapped"
objects  that may arise in them.  It is not argued that these
gravity models would explain  all the range of gravity
experiments\cite{nunez}. Probably, as in string
theory\cite{pimenpola,pimenpolb,wohlfarth}, higher order
perturbative corrections to the Einstein-Hilbert action
depending also on combinations of the
Riemann and Ricci tensors should also be present in a more
realistic theory, and the models considered here would be,
possibly, a limit of these theories.

To set the stage for our analysis, we begin by recalling some
properties of the NLG and EKG systems. For the former, we write
\begin{equation}
{\cal L}=\sqrt{-g} \, f(R) \, , \label{nlglagrangian}
\end{equation}
where $f$ is an arbitrary function of the scalar curvature $R$.
The field equations derived from~(\ref{nlglagrangian}) are
\begin{equation}
f'(R) R_{\mu \nu} -\frac{1}{2}g_{\mu \nu}f(R) +
 \left( g_{\mu \nu} \Box - \nabla_\mu \nabla_\nu \right) f'(R) =0 \, ,
\label{nlgeqs}
\end{equation}
in which $f'(R)\equiv \frac{df}{dR}\neq 0$.

Following a well--known
procedure\cite{potentialsa,potentialsb,potentialsd}, we now establish
the connection between this NLG theory and the EKG formalism. We first
define the new variables
\begin{equation}
e^{\sqrt{(2/3)\kappa_0}\Phi} = f'(R) \, ,
~\tilde{g}_{\mu \nu} = f'(R) g_{\mu \nu} \, , \label{nlgtransf}
\end{equation}
where the conformal transformation between the metrics is
invertable and then well defined. Inserting
this into Eq.~(\ref{nlgeqs}), we directly obtain the EKG equations
\begin{subequations}
\label{ekg}
\begin{eqnarray}
\tilde{G}_{\mu \nu} := \tilde{R}_{\mu \nu} -
\frac{1}{2}\tilde{g}_{\mu \nu} \tilde{R} &=&
\kappa_0 T_{\mu \nu} (\Phi) \, , \label{ekg1} \\
\tilde{\Box} \Phi -\frac{dV}{d\Phi} &=& 0 \, , \label{ekg2}
\end{eqnarray}
\end{subequations}
where
\begin{equation}
T_{\mu \nu} (\Phi) = \Phi_{,\mu} \Phi_{,\nu} -
\frac{1}{2} \tilde{g}_{\mu \nu} \left[ \Phi^{,\sigma} \Phi_{,\sigma}
+ 2 V(\Phi) \right] \, ,
\end{equation}
is the energy--momentum tensor of the scalar field $\Phi$ endowed with a scalar field potential $V(\Phi)$. The scalar field potential is related to $f$ and $R$ through
\begin{equation}
V(\Phi) = \frac{1}{2 \kappa_0 {f'}^2} \left( R f'- f \right) \, .
\label{nlgvphi}
\end{equation}
Also, $\kappa_0 = 8\pi G$ (we use units in which $\hbar=c=1$,
and then the Plank mass is $m_\textrm{Pl}=G^{1/2}$) and
$\tilde{\Box}=(1/\sqrt{-\tilde{g}})\partial_\mu
(\sqrt{-\tilde{g}}\tilde{g}^{\mu \nu} \partial_\nu)$ is the covariant
d'Alembertian operator. Notice that all metric quantities in the EKG system
will be denoted with a tilde. The corresponding Lagrangian is
\begin{equation}
{\cal \tilde{L}}=\sqrt{-\tilde{g}} \left[ \tilde{R} -
\Phi^{,\sigma} \Phi_{,\sigma} - 2 V(\Phi) \right] \, , \label{elagrangian}
\end{equation}
where we see that the scalar field is minimally coupled to the metric
tensor $\tilde{g}_{\mu \nu}$.

As a final point in this section, we would like to mention the
possible \textit{vacuum} solutions of Eqs.~(\ref{nlgeqs}) known so
far. It has been shown in\cite{mignemi} that, provided $f(R)=R+a_2 R^2
+ \cdots $\footnote{If $f''(R)\neq 0$, then the
  Lagrangian~(\ref{nlglagrangian}) is called
  \textit{$R$--regular}\cite{potentialsd}.}, if $a_2>0$ then the only
static spherically symmetric and asymptotically flat solution with a
regular horizon is the Schwarzschild solution. As it can be seen, the
statement above is not as strong as the Birkhoff theorem that arises
in General Relativity, that a spherically symmetric gravitational
field in empty space must be static, with a metric given by the
Schwarzschild one\cite{gravitation,grweinberg}.

We should emphasize that, as we shall show below, the
transformation~(\ref{nlgtransf}) will allow us to map known EKG
oscillaton solutions onto a \textit{full} NLG theory; that is, the
transformation will not be performed on the perturbative expansion of
a NLG theory.

Although we will exhibit the corresponding perturbative expansion
of the resulting NLG Lagrangian, it is not the aim of this work to find
solutions to any NLG perturbative expansion. Nevertheless, the
search for oscillaton kind of solutions to a NLG theory could be of
interest for the case of other Lagrangians that, in particular, are expressed
through a (well defined) perturbative expansion.

Even though we have shown the \textit{formal} equivalence between two
conformally related frames, it should be reminded that this does not
imply a \textit{physical} equivalence too. We will make use of the
former to find solutions to Eq.~(\ref{nlgeqs}), and will comment on
the latter equivalence in the last sections. However, the present manuscript
should not be seen as a discussion on the trueness of one particular
frame, but rather as another example that may help us to understand
some of the (yet hidden) particularities of the NLG-frame.

The sections are organized as follows. In section \ref{oscillatons},
we shall present the solutions to the EKG system and many of their interesting
features. In section \ref{massradius}, we find the corresponding NLG
theory and exhibit a way to compare the mass--radius relation of the
objects formed in both systems. Section \ref{finale} is devoted to
final remarks.

\section{Oscillatons: numerical solutions of the EKG system}
\label{oscillatons}
Regular and asymptotically flat solutions of the EKG
equations~(\ref{ekg}) are called oscillatons, and then we expect them
to be related to the vacuum solutions of the field
equations~(\ref{nlgeqs}). As an example, we will focus our attention
in the simplest oscillaton, which arises from a quadratic scalar
potential of the form $V(\Phi)=(1/2)m^2_\Phi \Phi^2$, in the
spherically symmetric case. Using the polar-areal slicing, we write
the line element as
\begin{equation}
ds^2 = -\tilde{\alpha}^2(\tilde{t},\tilde{r}) d\tilde{t}^2 +
\tilde{a}^2(\tilde{t},\tilde{r}) d\tilde{r}^2 + \tilde{r}^2 d\Omega \,
, \label{efmetric}
\end{equation}
where $\tilde{\alpha} (\tilde{t},\tilde{r})$ and $\tilde{a}
(\tilde{t},\tilde{r})$ are the metric functions that appear on the left hand side of Eq.~(\ref{ekg1}).

It is not possible to find analytical solutions to the EKG
system, but Eqs.~(\ref{ekg}) can be solved numerically instead.
In order to find well behaved solutions, the numerical solution
should be fully time-dependent. The most popular solution is
due to Seidel \& Suen \cite{seidel91,seidel94}, that used
Fourier expansions of the functions involved in the EKG equations.
The method has been refined in \cite{laul2002a,laul2002b} to
facilitate the numerical solution; and it is the latter
which is briefly described next to construct the oscillaton solutions.

The Fourier series of the fields are of the form
\begin{subequations}
\label{expansions}
\begin{eqnarray}
\sqrt{8\pi G} \, \Phi(\tilde{t},\tilde{r}) &=& \sum^{\infty}_{j=1}
\phi_{j}(\tilde{r}) \cos(j\omega \tilde{t}) \, , \label{expansionsa} \\
\tilde{\alpha}^2 (\tilde{t},\tilde{r}) &=& \sum^{\infty}_{j=0}
\tilde{\alpha}_j (\tilde{r}) \cos(j\omega \tilde{t}) \,,\label{expansionsb} \\
\tilde{a}^2 (\tilde{t},\tilde{r}) &=& \sum^{\infty}_{j=0} \tilde{a}_j(\tilde{r}) \cos(j\omega \tilde{t}) \, ,
\label{expansionsc}
\end{eqnarray}
\end{subequations}
where $\omega$ is the fundamental angular frequency of the system.

Imposing boundary conditions of regularity at the origin
and of asymptotic flatness, the EKG system becomes an eigenvalue
problem. For each value of, say, $\phi_1(0)$, there is a set of
eigenvalues of $\omega$ and $\tilde{\alpha}_j(0)$ for which the
boundary conditions are fulfilled (for more details see \cite{laul2002b}).
The Fourier series have to be truncated by hand, and just a few
of the Fourier coefficients are taken into account. However,
all the solutions show convergence: the higher the Fourier mode,
the less it contributes to the series.

\begin{figure}
\includegraphics[width=8cm]{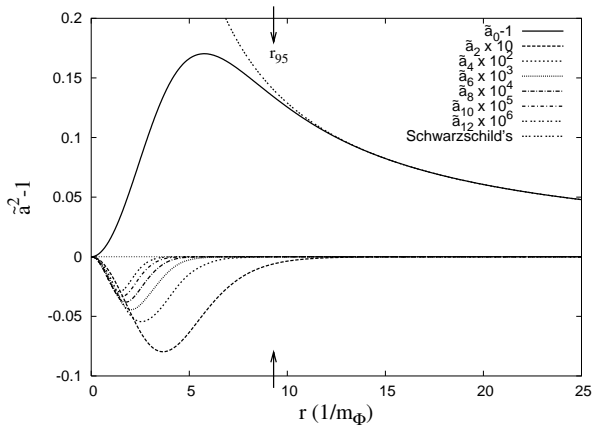}
\caption{\label{fig:nlg1} The Fourier coeffiecients
corresponding to the radial metric function $\tilde g_{rr}=\tilde a^2$
according to expansion~(\ref{expansionsc}), for an oscillaton with
total mass $M=0.571 (m^2_\textrm{Pl}/m_\Phi)$ and for which
$\phi_1(0)=2\sqrt{2}$; also shown is its $95$\% radius $r_{95}=9.34m^{-1}_\Phi$ (For oscillatons, $m_\Phi$ becomes the natural unit of distance).
The Fourier series was truncated at $j=12$, and the profiles
were appropriately scaled to show them all in the same plot.
The convergence of the solution is manifest.}
\end{figure}

Typical profiles of the Fourier modes of the radial metric field
$\tilde{g}_{rr}\equiv \tilde{a}^2$ are shown in Fig.~\ref{fig:nlg1},
up to $j=12$, which corresponds to an oscillaton with total mass
 $M_\textrm{osc}=0.571 (m^2_\textrm{Pl}/m_\Phi)$ (here $\phi_1(0)=2\sqrt{2}$).
It should be noticed that the time-dependent Fourier terms
are confined, and that outside a typical radius, say
  $\tilde{r}_{95}$ (the radius within which $95$\% of the total mass
  is contained in), the metric matches the Schwarzschild solution,
\begin{equation}
\tilde{\alpha}^2 (\tilde{t},\tilde{r}>\tilde{r}_{95}) = \tilde{a}^{-2}(\tilde{t},\tilde{r}>\tilde{r}_{95})
= 1-\frac{2GM_\textrm{osc}}{\tilde{r}} \, . \label{outmetric}
\end{equation}
We should add here that the mass of an
oscillaton is finite due to the fact that the scalar field
vanishes exponentially for $\tilde{r} > \tilde{r}_{95}$.

A curious point is that, by construction of the solution, $\Phi$ has
only odd multiples of the fundamental angular frequency $\omega$, and
the metric functions $\tilde{\alpha}$ and $\tilde{a}$ have the even
ones. This seems to indicate that the oscillaton in
Fig.~\ref{fig:nlg1} and alike are the simplest configurations (the
less massive) one can construct for the EKG system: the inclusion of
all cosine coefficients or of sine terms would result in more massive
oscillatons\footnote{The mass of an oscillaton is proportional to
  the energy density $\rho_\Phi=(1/2) (\alpha^{-2} \dot{\Phi}^2 + a^{-2}
  \Phi^{\prime 2}+m^2_\Phi \Phi^2)$, and then a larger number of
  Fourier coefficients in~(\ref{expansionsa}) means a larger mass of
  the scalar configuration.}. 

An important issue of oscillatons that have not been fully
tackled is that of their stability. The numerical experience
so far points out that there exist intrinsically \textit{stable}
(S-branch) and \textit{unstable} (U-branch) oscillatons \cite{phi2003};
cf.~the more detailed description in
catastrophe language in \cite{KMS91a,KMS91b}. The U-branch solutions migrate
away from the equilibrium solution under small perturbations.
Stable oscillatons, on the other hand, do not migrate if
perturbed a little. Moreover, it seems that they play the
role of \textit{final} states an arbitrary scalar field
configuration evolves towards to. However, there is not a
definite proof for the stability of oscillatons, but it
seems that, if oscillatons are not fully stable, they are
long lived at least\cite{seidel91,phi2003,don2003}.

\section{The mass--radius relation in the NLG and EKG systems}
\label{massradius}
Now, we are to write explicitly the connection between the
NLG and the EKG systems for the particular example of a quadratic oscillaton.
Eq.~(\ref{nlgvphi}) is usually taken as a differential equation
for $f(R)$ to be solved in terms of the curvature $R$. As can be seen
from Eq.~(\ref{nlgtransf}), one obtains a highly non-linear
differential equation, which is very difficult to solve.

This can be ameliorated if we derive Eq.~(\ref{nlgvphi})
once again with respect to $R$, which yields
\begin{equation}
2\kappa_0 f'' \left[ 2f'V +\frac{1}{\sqrt{(2/3)\kappa_0}}
f' \frac{dV}{d\Phi} - \frac{R}{2\kappa_0}\right] = 0 \, . \label{nlgvphi2}
\end{equation}
The obvious and trivial solution, which appears for all
cases, is $f''(R)=0$, which means $f(R)=AR+B$, with $A$
and $B$ arbitrary constants, and also that
$\Phi=\textrm{const.}$\footnote{Notice that Eq.~(\ref{nlgtransf}) requires
  $f^\prime \neq 0$ to have a non-trivial transformation, while
  Eq.~(\ref{nlgvphi2}) needs $f^{\prime \prime} \neq 0$ to have a
  non-trivial solution for $f(R)$.}.

It turns out that, in the general case, $f$ and $R$ can be
given in parametric form in terms of $\Phi$ by means of
Eqs.~(\ref{nlgtransf}),~(\ref{nlgvphi}) and~(\ref{nlgvphi2}). To simplify
the calculations, we define a new dimensionless variable by
$\ln(x)=\sqrt{(2/3)\kappa_0}\Phi$. Thus,
\begin{subequations}
\label{parametric}
\begin{eqnarray}
f &=& 2\kappa_0 x^2 \left[ V(x) + x \frac{dV(x)}{dx} \right] \, ,
\label{parametrica} \\
R &=& 4\kappa_0 x \left[ V(x) + \frac{x}{2} \frac{dV(x)}{dx} \right] \, .
\label{parametricb}
\end{eqnarray}
\end{subequations}
In this form, it is easily seen that the Liouville theory,
for which $V(\Phi)=V_0 e^{\lambda \Phi}$, where $V_0$ and $\lambda$
are constants, is one of the scalar-tensor cases that can be solved
exactly\cite{potentialsa,potentialsb,potentialsd}. However, it seems
that the Liouville theory does not provide regular nor asymptotically
flat solutions of the EKG system, see for
instance\cite{matos2000a}\footnote{In case of a complex scalar field,
  a $U(1)$-Liouville potential can provide of stable boson
  stars\cite{ST00}}.

On the other hand, for the quadratic case $V(\Phi)=(1/2)m^2_\Phi \Phi^2$, Eqs.~(\ref{parametric}) are rewritten as \cite{potentialsd}
\begin{subequations}
\label{parasquare}
\begin{eqnarray}
f &=& \frac{3}{2}m^2_\Phi x^2 \ln(x) \left[ \ln(x)+2 \right] \, ,
\label{parasquarea} \\
R &=& 3 m^2_\Phi x \ln(x) \left[ \ln(x)+1\right] \, . \label{parasquareb}
\end{eqnarray}
\end{subequations}
A Taylor expansion of Eqs.~(\ref{parasquare}) around $f(0)=0$ ($x=1$)
gives 
\begin{equation}
f(R)=R+\frac{1}{6m^2_\Phi}R^2 -\frac{1}{18m^4_\Phi}R^3 + \mathcal{O}(R^4) \, .
\label{nlgseries}
\end{equation}
Hence, a quadratic oscillaton corresponds to an $R$-regular NLG
theory.

It has been argued before that Lagrangians of the form~(\ref{nlgseries}) accomplish some desirable features\cite{sokol95}. For instance, the presence of an $R$-term ensures that the EKG exists near Minkowski spacetime, and the existence of an $a_2 R^2$-term ensures regularity of the conformal transformation to flat space; moreover, if $a_2 >0$ then the Minkowski space is a stable ground state solution. It also seems that curvature corrections at all orders are essential in order to regulate gravity\cite{wohlfarth}.

It should be remarked that NLG theories with characteristics as
mentioned in the above paragraph correspond to a very particular class
of scalar potentials and probably are a limit of a more general
theory\cite{pimenpola}. From Eqs.~(\ref{parasquare}), it can be seen
that Lagrangians of the form~(\ref{nlgseries}) correspond to the case
in which $V(0)=\frac{dV}{d\Phi}(0)=0$, i.e., the scalar potential
$V(\Phi)$ has a critical point at $\Phi=0$. Explicitly, the first
Taylor coefficients for this type of potentials, calculated from
Eqs.~(\ref{parametric}), are
\begin{eqnarray}
a_0 &=& 0 \, , \nonumber \\
a_1 &=& \left. x \right|_{x=1} = 1 \, , \nonumber \\
a_2 &=& \frac{1}{2} \left. \frac{1}{\dot{R}} \right|_{x=1} =
\frac{1}{6V^{\prime \prime}(0)} \, , \nonumber \\
a_3 &=& -\frac{1}{6} \left. \frac{\ddot{R}}{\dot{R}^3} \right|_{x=1} =
-\frac{1}{18 \left[ V^{\prime \prime} (0) \right]^2} \left(
1+\frac{1}{\sqrt{6\kappa_0}} \frac{V^{\prime \prime
    \prime}(0)}{V^{\prime \prime}(0)} \right) \, , \nonumber \\
&\cdots& \nonumber
\end{eqnarray}
where $\dot{}\equiv \frac{d}{dx}$ and ${}^\prime \equiv
  \frac{d}{d\Phi}$. Higher Fourier terms depend on higher derivatives
  of the scalar potential in a complicated way, so that we do not show
  them here. However, it is easily seen that \emph{$a_2 >0$ only if
  the scalar field is also massive, i.e., $m^2_\Phi \equiv V^{\prime
  \prime}(0)>0$}.


What else can we say of the NLG theory in~(\ref{parasquare})? First,
the line element in the NLG theory is spherically symmetric and fully
time-dependent, properties preserved by the conformal
transformation~(\ref{nlgtransf}). The metric fields have Fourier
expansions of the form
\begin{subequations}
\label{nlgexpansions}
\begin{eqnarray}
\alpha^2 &=& e^{-\sqrt{(2/3)\kappa_0}\Phi} \tilde{\alpha}^2 = \sum^{\infty}_{j=0} \alpha_j (\tilde{r}) \cos(j\omega \tilde{t}) \, ,
\label{nlgexpansionsa} \\
a^2 &=& e^{-\sqrt{(2/3)\kappa_0}\Phi} \tilde{a}^2 = \sum^{\infty}_{j=0} a_j(\tilde{r}) \cos(j\omega \tilde{t}) \, ,
\label{nlgexpansionsb}
\end{eqnarray}
\end{subequations}
where, in contrast to Eqs.~(\ref{expansions}) of an EKG-oscillaton,
all the Fourier coefficients, odd and even, are non-trivial. Here, it
is the scalar field $\Phi$ which provides the odd coefficients.

Second, the \textit{exterior} form for both metrics $g$ and
$\tilde{g}$ is the Schwarzschild solution~(\ref{outmetric}). In
consequence, the ADM mass is the same in both the NLG and the EKG.

Therefore, this theory permits the existence of regular
objects made of \textit{pure} gravity, which are, in addition,
\textit{vacuum} solutions. Hence, there may be in some cases a
  complementary second part of the theorem mentioned
  before\cite{mignemi}. We have shown here that there is a non-linear
  Lagrangian of the form $f(R)=R+a_2 R^2 + \cdots $, with $a_2>0$,
  that has \emph{two type of asymptotically flat vacuum solutions}:
  Schwarzschild (static), and that corresponding to oscillatons
  (time-dependent). It would be interesting to investigate whether the
  above statement is true for any massive real scalar field, since it
  has not been proved completely that the EKG equations have regular
  stable solutions only in the case $m^2_\Phi > 0$; though this seems
  to be the case, see for instance\cite{ulises}.

  We observe that the form of \textit{all} of the coefficients in the
  Taylor expansion of the exact result~(\ref{parasquare}) is
  determinant for the existence of oscillaton solutions; had we started with the approximate expansion~(\ref{nlgseries}), the results
  could have in general differed from the results obtained in here, even
  approximately.

On the other hand, it should be noticed that the simplest solution to
Eqs.~(\ref{nlgeqs}) is the metric $g_{\mu \nu}$ whose coefficientes
are expanded in a Fourier series of the
form~(\ref{nlgexpansions}). As we said before, this metric is
spherically symmetric but is not written in its so--called standard
form. An observer in the NLG frame would, however, notice that the
coordinates $\tilde{t},\tilde{r}$ are its usual Schwarzschild
coordinates far from $\tilde{r}=0$. This would indicate her/him that
there is an \textit{exterior} (static) vacuum solution and an
\textit{interior} (time-dependent) one.

\begin{figure}
\includegraphics[width=8cm]{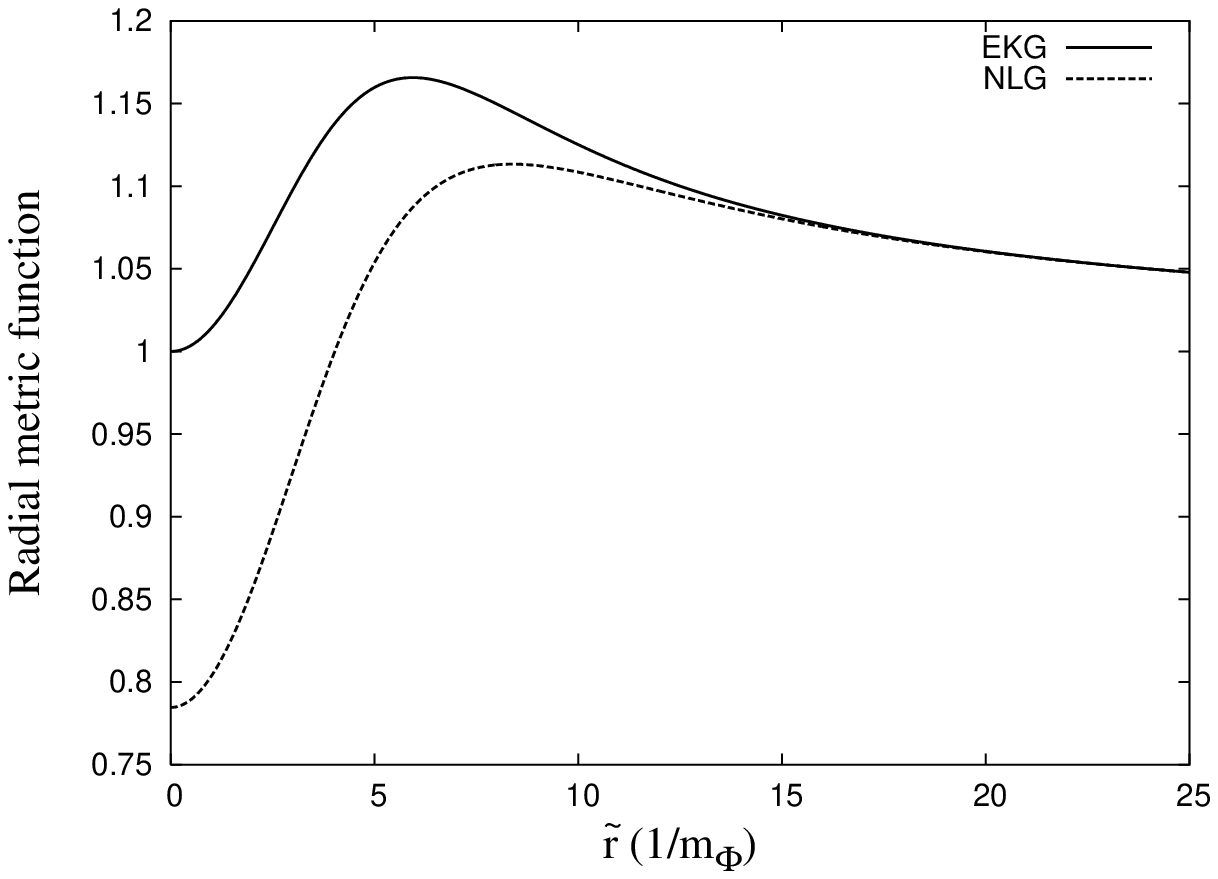}
\includegraphics[width=8cm]{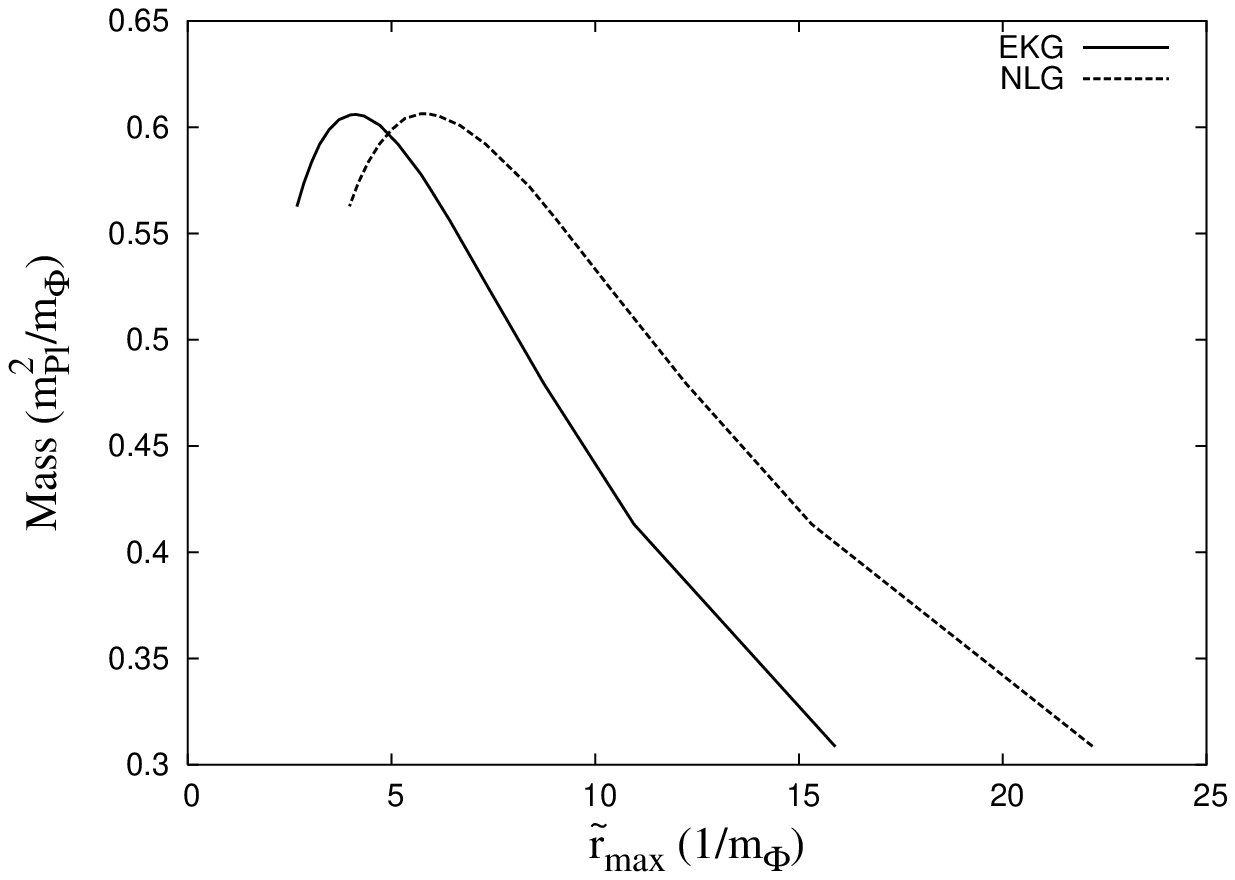}
\caption{\label{fig:nlg2} (Top) Comparison of the radial metric function
 obtained from the EKG and the NLG cases for the same case shown in
 Fig.~\ref{fig:nlg1}. It can be seen that the NLG solution is
 asymptotically flat and coincides with the EKG solution for
 $\tilde{r} \rightarrow \infty$; but it is conformally flat at
 $\tilde{r}=0$. (Bottom) $M$ vs $\tilde{r}_\textrm{max}$ plots for the
 EKG and NLG solutions, the configuration corresponding to the
 leftmost (rightmost) point of the plot has
 $\sqrt{\kappa_0}\Phi(0,0)=0.8$ ($0.01$). The NLG-oscillatons are less
 compact that their EKG-counterparts, so that they are distinguishable
 in principle, see text for details.}
\end{figure}

This observer would also find more handy to continue using
$(\tilde{t},\tilde{r})$ as her/his metric coordinates, and then to preserve $g_{\mu \nu}$ in its simplest form. Thinking of this possibility, we show a
comparison between the radial metric functions in the EKG and NLG
cases in Fig.~\ref{fig:nlg2}. We also plot the corresponding $M$
vs $\tilde{r}_{max}$ graphs for both cases, where $\tilde{r}_\textrm{max}$ is
the position of the maximum of the radial metric
function~(\ref{nlgexpansionsb}) at $\tilde{t}=0$. There is a maximum
mass for the NLG case, which could be interpreted as an indication of
the existence of Stable and Unstable branches, as in the case of the
EKG theory.

\section{Final remarks}
\label{finale}
If the NLG theory in~(\ref{parasquare}) were the fundamental gravity
theory, there would be many \textit{dark pure-gravity} objects,
corresponding to oscillatons, in the universe. For a given mass, the
object in the NLG theory is bigger than its EKG counter part
(Fig.~\ref{fig:nlg2}), so that one could, in principle, discriminate
between them. Ultimately, by these means we could determine which
system, the NLG or the EKG, is being measured.

On the other hand, we could ask at this point: how can these objects
be formed from the NLG point of view? As we mentioned before, in the
EKG, a self-gravitating scalar field evolves and settle-down onto an
S-oscillaton. Because of the spherical symmetry of the system, the
only mechanism of relaxation permitted is the emission of
\textit{scalar radiation} (scalar field matter). This process has been
dubbed \textit{gravitational
  cooling}\cite{seidel94,phi2003,laul2003a,laul2004a}, which has been
shown to be a very efficient mechanism for the relaxation of
self--gravitating scalar fields.

On the other hand, any initial scalar field configuration in the EKG
frame can be conformally transformed into a pure gravity configuration
in the NLG. Moreover, also the complete evolution of the EKG system
can be conformally transformed and followed in the NLG. As the system
is spherically symmetric also in the NLG, gravitational radiation is
forbidden, and so there should exist a mechanism for the relaxation
which would be the ``conformal'' partner of the gravitational cooling.

The only explanation we foresee is that, in the higher derivative NLG
frame, gravity must be allowed to have a spin-$0$ component, which
would provide the channel for gravitational cooling. In fact, the existence of \textit{scalar gravitational waves} has been studied before, together with the idea that actual interferometers built for the detection of gravitational waves can also detect a scalar component of gravitational radiation\cite{belluci,maggiore}.

It would be interesting to determine the scalar gravitational
waves emmited for the system discussed here, for which one would
expect that the non-linear-terms would, in a certain manner, take the
role of the scalar field. The procedure would be the full numerical
evolution of Eqs.~(\ref{nlgeqs}). For this, it would also help the formal
equivalence between the NLG and the EKG, as the numerical evolution
of the metric fields $\alpha$ and $a$ can be mapped directly
from that of $\tilde{\alpha}$, $\tilde{a}$ and $\Phi$. These
calculations are in progress and will be reported elsewhere.

\begin{acknowledgments}
We acknowledge Ricardo Becerril for sharing the computer code to
generate numerical solutions for oscillatons. This work is partially
supported by a CONACYT grant Number 37851-E, PROMEP UGTO-CA-3 and
projects from Universidad de Guanajuato.
\end{acknowledgments}

\bibliography{nlgphi}

\begin{thebibliography}{42}
\expandafter\ifx\csname natexlab\endcsname\relax\def\natexlab#1{#1}\fi
\expandafter\ifx\csname bibnamefont\endcsname\relax
  \def\bibnamefont#1{#1}\fi
\expandafter\ifx\csname bibfnamefont\endcsname\relax
  \def\bibfnamefont#1{#1}\fi
\expandafter\ifx\csname citenamefont\endcsname\relax
  \def\citenamefont#1{#1}\fi
\expandafter\ifx\csname url\endcsname\relax
  \def\url#1{\texttt{#1}}\fi
\expandafter\ifx\csname urlprefix\endcsname\relax\def\urlprefix{URL }\fi
\providecommand{\bibinfo}[2]{#2}
\providecommand{\eprint}[2][]{\url{#2}}

\bibitem[{\citenamefont{Lanczos}(1938)}]{lanczos}
\bibinfo{author}{\bibfnamefont{C.}~\bibnamefont{Lanczos}},
  \bibinfo{journal}{Ann.\ Math.} \textbf{\bibinfo{volume}{39}},
  \bibinfo{pages}{842} (\bibinfo{year}{1938}).

\bibitem[{\citenamefont{Flanagan}(2004{\natexlab{a}})}]{flanagana}
\bibinfo{author}{\bibfnamefont{{\'Eanna}.~{\'E}.} \bibnamefont{Flanagan}},
  \bibinfo{journal}{Class. \ Quantum \ Grav.} \textbf{\bibinfo{volume}{21}},
  \bibinfo{pages}{3817} (\bibinfo{year}{2004}{\natexlab{a}}),
  \eprint{gr-qc/0403063}.

\bibitem[{\citenamefont{Flanagan}(2004{\natexlab{b}})}]{flanaganb}
\bibinfo{author}{\bibfnamefont{{\'Eanna}.~{\'E}.} \bibnamefont{Flanagan}},
  \bibinfo{journal}{Class. \ Quantum \ Grav.} \textbf{\bibinfo{volume}{21}},
  \bibinfo{pages}{417} (\bibinfo{year}{2004}{\natexlab{b}}),
  \eprint{gr-qc/0309015}.

\bibitem[{\citenamefont{Utiyama and DeWitt}(1962)}]{pimentela}
\bibinfo{author}{\bibfnamefont{R.}~\bibnamefont{Utiyama}} \bibnamefont{and}
  \bibinfo{author}{\bibfnamefont{B.}~\bibnamefont{DeWitt}},
  \bibinfo{journal}{J. \ Math. \ Phys.} \textbf{\bibinfo{volume}{3}},
  \bibinfo{pages}{608} (\bibinfo{year}{1962}).

\bibitem[{\citenamefont{Stelle}(1977)}]{pimentelb}
\bibinfo{author}{\bibfnamefont{K.}~\bibnamefont{Stelle}},
  \bibinfo{journal}{Phys. \ Rev. \ D} \textbf{\bibinfo{volume}{16}},
  \bibinfo{pages}{953} (\bibinfo{year}{1977}).

\bibitem[{\citenamefont{Strominger}(1984)}]{pimentelc}
\bibinfo{author}{\bibfnamefont{A.}~\bibnamefont{Strominger}},
  \bibinfo{journal}{Phys. \ Rev. \ D} \textbf{\bibinfo{volume}{30}},
  \bibinfo{pages}{2257} (\bibinfo{year}{1984}).

\bibitem[{\citenamefont{Chern and Simons}(1974)}]{pimentel2a}
\bibinfo{author}{\bibfnamefont{S.}~\bibnamefont{Chern}} \bibnamefont{and}
  \bibinfo{author}{\bibfnamefont{J.}~\bibnamefont{Simons}},
  \bibinfo{journal}{Ann. \ Math.} \textbf{\bibinfo{volume}{99}},
  \bibinfo{pages}{48} (\bibinfo{year}{1974}).

\bibitem[{\citenamefont{Lovelock}(1971)}]{pimentel2b}
\bibinfo{author}{\bibfnamefont{D.}~\bibnamefont{Lovelock}},
  \bibinfo{journal}{J. \ Math. \ Phys.} \textbf{\bibinfo{volume}{12}},
  \bibinfo{pages}{498} (\bibinfo{year}{1971}).

\bibitem[{\citenamefont{Mena-Marug{\'a}n}(1990)}]{pimentel2c}
\bibinfo{author}{\bibfnamefont{G.~A.} \bibnamefont{Mena-Marug{\'a}n}},
  \bibinfo{journal}{Phys. \ Rev. \ D} \textbf{\bibinfo{volume}{42}},
  \bibinfo{pages}{2607} (\bibinfo{year}{1990}).

\bibitem[{\citenamefont{Ban{\~a}dos et~al.}(1992)\citenamefont{Ban{\~a}dos,
  Teitelboim, and Zanelli}}]{pimentel2d}
\bibinfo{author}{\bibfnamefont{M.}~\bibnamefont{Ban{\~a}dos}},
  \bibinfo{author}{\bibfnamefont{C.}~\bibnamefont{Teitelboim}},
  \bibnamefont{and} \bibinfo{author}{\bibfnamefont{J.}~\bibnamefont{Zanelli}},
  \bibinfo{journal}{Phys. \ Rev. \ Lett.} \textbf{\bibinfo{volume}{69}},
  \bibinfo{pages}{184} (\bibinfo{year}{1992}).

\bibitem[{\citenamefont{Ban{\~a}dos et~al.}(1993)\citenamefont{Ban{\~a}dos,
  Henneaux, Teitelboim, and Zanelli}}]{pimentel2e}
\bibinfo{author}{\bibfnamefont{M.}~\bibnamefont{Ban{\~a}dos}},
  \bibinfo{author}{\bibfnamefont{M.}~\bibnamefont{Henneaux}},
  \bibinfo{author}{\bibfnamefont{C.}~\bibnamefont{Teitelboim}},
  \bibnamefont{and} \bibinfo{author}{\bibfnamefont{J.}~\bibnamefont{Zanelli}},
  \bibinfo{journal}{Phys. \ Rev. \ D} \textbf{\bibinfo{volume}{48}},
  \bibinfo{pages}{2797} (\bibinfo{year}{1993}).

\bibitem[{\citenamefont{Deser and Redlich}(1986)}]{pimenpola}
\bibinfo{author}{\bibfnamefont{S.}~\bibnamefont{Deser}} \bibnamefont{and}
  \bibinfo{author}{\bibfnamefont{N.}~\bibnamefont{Redlich}},
  \bibinfo{journal}{Phys. Lett. B} \textbf{\bibinfo{volume}{176}},
  \bibinfo{pages}{350} (\bibinfo{year}{1986}).

\bibitem[{\citenamefont{Polchinski}(1998)}]{pimenpolb}
\bibinfo{author}{\bibfnamefont{J.}~\bibnamefont{Polchinski}},
  \emph{\bibinfo{title}{String Theory}}, vol. \bibinfo{volume}{I, II}
  (\bibinfo{publisher}{Cambridge University Press}, \bibinfo{year}{1998}).

\bibitem[{\citenamefont{Mielke et~al.}(1997)\citenamefont{Mielke, Obreg{\'o}n,
  and Mac{\'i}as}}]{potentialsa}
\bibinfo{author}{\bibfnamefont{E.~W.} \bibnamefont{Mielke}},
  \bibinfo{author}{\bibfnamefont{O.}~\bibnamefont{Obreg{\'o}n}},
  \bibnamefont{and}
  \bibinfo{author}{\bibfnamefont{A.}~\bibnamefont{Mac{\'i}as}},
  \bibinfo{journal}{Phys. Lett. B} \textbf{\bibinfo{volume}{391}},
  \bibinfo{pages}{281} (\bibinfo{year}{1997}).

\bibitem[{\citenamefont{Ben{\'i}tez et~al.}(1997)\citenamefont{Ben{\'i}tez,
  Mac{\'i}as, Mielke, Obreg{\'o}n, and Villanueva}}]{potentialsb}
\bibinfo{author}{\bibfnamefont{J.}~\bibnamefont{Ben{\'i}tez}},
  \bibinfo{author}{\bibfnamefont{A.}~\bibnamefont{Mac{\'i}as}},
  \bibinfo{author}{\bibfnamefont{E.~W.} \bibnamefont{Mielke}},
  \bibinfo{author}{\bibfnamefont{O.}~\bibnamefont{Obreg{\'o}n}},
  \bibnamefont{and} \bibinfo{author}{\bibfnamefont{V.~M.}
  \bibnamefont{Villanueva}}, \bibinfo{journal}{Int. \ J. \ Mod. \ Phys. \ A}
  \textbf{\bibinfo{volume}{12}}, \bibinfo{pages}{2835} (\bibinfo{year}{1997}).

\bibitem[{\citenamefont{Mijic et~al.}(1986)\citenamefont{Mijic, Morris, and
  Suen}}]{potentialsc}
\bibinfo{author}{\bibfnamefont{M.~B.} \bibnamefont{Mijic}},
  \bibinfo{author}{\bibfnamefont{M.~S.} \bibnamefont{Morris}},
  \bibnamefont{and} \bibinfo{author}{\bibfnamefont{W.}~\bibnamefont{Suen}},
  \bibinfo{journal}{Phys. \ Rev. \ D} \textbf{\bibinfo{volume}{34}},
  \bibinfo{pages}{2934} (\bibinfo{year}{1986}).

\bibitem[{\citenamefont{Starobinsky}(1980)}]{potentialsf}
\bibinfo{author}{\bibfnamefont{A.~A.} \bibnamefont{Starobinsky}},
  \bibinfo{journal}{Phys. Lett. B} \textbf{\bibinfo{volume}{91}},
  \bibinfo{pages}{99} (\bibinfo{year}{1980}).

\bibitem[{\citenamefont{Kalara et~al.}(1990)\citenamefont{Kalara, Kaloper, and
  Olive}}]{potentialsg}
\bibinfo{author}{\bibfnamefont{S.}~\bibnamefont{Kalara}},
  \bibinfo{author}{\bibfnamefont{N.}~\bibnamefont{Kaloper}}, \bibnamefont{and}
  \bibinfo{author}{\bibfnamefont{K.}~\bibnamefont{Olive}},
  \bibinfo{journal}{Nucl. \ Phys. \ B} \textbf{\bibinfo{volume}{341}},
  \bibinfo{pages}{252} (\bibinfo{year}{1990}).

\bibitem[{\citenamefont{Herrera et~al.}(1995)\citenamefont{Herrera, Contreras,
  and del Campo}}]{potentialsh}
\bibinfo{author}{\bibfnamefont{R.}~\bibnamefont{Herrera}},
  \bibinfo{author}{\bibfnamefont{C.}~\bibnamefont{Contreras}},
  \bibnamefont{and} \bibinfo{author}{\bibfnamefont{S.}~\bibnamefont{del
  Campo}}, \bibinfo{journal}{Class. \ Quantum \ Grav.}
  \textbf{\bibinfo{volume}{12}}, \bibinfo{pages}{1937} (\bibinfo{year}{1995}).

\bibitem[{\citenamefont{Seidel and Suen}(1991)}]{seidel91}
\bibinfo{author}{\bibfnamefont{E.}~\bibnamefont{Seidel}} \bibnamefont{and}
  \bibinfo{author}{\bibfnamefont{W.-M.} \bibnamefont{Suen}},
  \bibinfo{journal}{Phys. \ Rev. \ Lett.} \textbf{\bibinfo{volume}{66}},
  \bibinfo{pages}{1659} (\bibinfo{year}{1991}).

\bibitem[{\citenamefont{Seidel and Suen}(1994)}]{seidel94}
\bibinfo{author}{\bibfnamefont{E.}~\bibnamefont{Seidel}} \bibnamefont{and}
  \bibinfo{author}{\bibfnamefont{W.-M.} \bibnamefont{Suen}},
  \bibinfo{journal}{Phys. \ Rev. \ Lett.} \textbf{\bibinfo{volume}{72}},
  \bibinfo{pages}{2516} (\bibinfo{year}{1994}).

\bibitem[{\citenamefont{Ure{\~n}a-L{\'o}pez}(2002)}]{laul2002a}
\bibinfo{author}{\bibfnamefont{L.~A.} \bibnamefont{Ure{\~n}a-L{\'o}pez}},
  \bibinfo{journal}{Class. \ Quantum \ Grav.} \textbf{\bibinfo{volume}{19}},
  \bibinfo{pages}{2617} (\bibinfo{year}{2002}), \eprint{gr-qc/0104093}.

\bibitem[{\citenamefont{Ure{\~n}a-L{\'o}pez
  et~al.}(2002)\citenamefont{Ure{\~n}a-L{\'o}pez, Matos, and
  Becerril}}]{laul2002b}
\bibinfo{author}{\bibfnamefont{L.~A.} \bibnamefont{Ure{\~n}a-L{\'o}pez}},
  \bibinfo{author}{\bibfnamefont{T.}~\bibnamefont{Matos}}, \bibnamefont{and}
  \bibinfo{author}{\bibfnamefont{R.}~\bibnamefont{Becerril}},
  \bibinfo{journal}{Class. \ Quantum \ Grav.} \textbf{\bibinfo{volume}{19}},
  \bibinfo{pages}{6259} (\bibinfo{year}{2002}).

\bibitem[{\citenamefont{Alcubierre et~al.}(2003)\citenamefont{Alcubierre,
  Becerril, Guzm{\'a}n, Matos, N{\'u}{\~n}ez, and
  Ure{\~n}a-L{\'o}pez}}]{phi2003}
\bibinfo{author}{\bibfnamefont{M.}~\bibnamefont{Alcubierre}},
  \bibinfo{author}{\bibfnamefont{R.}~\bibnamefont{Becerril}},
  \bibinfo{author}{\bibfnamefont{F.~S.} \bibnamefont{Guzm{\'a}n}},
  \bibinfo{author}{\bibfnamefont{T.}~\bibnamefont{Matos}},
  \bibinfo{author}{\bibfnamefont{D.}~\bibnamefont{N{\'u}{\~n}ez}},
  \bibnamefont{and} \bibinfo{author}{\bibfnamefont{L.~A.}
  \bibnamefont{Ure{\~n}a-L{\'o}pez}}, \bibinfo{journal}{Class. \ Quantum \
  Grav.} \textbf{\bibinfo{volume}{20}}, \bibinfo{pages}{2883}
  (\bibinfo{year}{2003}), \eprint{gr-qc/0301105}.

\bibitem[{\citenamefont{Guzm{\'a}n and Ure{\~n}a-L{\'o}pez}(2003)}]{laul2003a}
\bibinfo{author}{\bibfnamefont{F.~S.} \bibnamefont{Guzm{\'a}n}}
  \bibnamefont{and} \bibinfo{author}{\bibfnamefont{L.~A.}
  \bibnamefont{Ure{\~n}a-L{\'o}pez}}, \bibinfo{journal}{Phys. \ Rev. \ D}
  \textbf{\bibinfo{volume}{68}}, \bibinfo{pages}{024023}
  (\bibinfo{year}{2003}), \eprint{astro-ph/0303440}.

\bibitem[{\citenamefont{Page}(2004)}]{don2003}
\bibinfo{author}{\bibfnamefont{D.}~\bibnamefont{Page}}, \bibinfo{journal}{Phys.
  \ Rev. \ D} \textbf{\bibinfo{volume}{70}}, \bibinfo{pages}{023002}
  (\bibinfo{year}{2004}), \eprint{gr-qc/0310006}.

\bibitem[{\citenamefont{N{\'u}{\~n}ez and Solganik}(2004)}]{nunez}
\bibinfo{author}{\bibfnamefont{A.}~\bibnamefont{N{\'u}{\~n}ez}}
  \bibnamefont{and} \bibinfo{author}{\bibfnamefont{S.}~\bibnamefont{Solganik}},
  \emph{\bibinfo{title}{The content of f(r) gravity}} (\bibinfo{year}{2004}),
  \eprint{hep-th/0403195}.

\bibitem[{\citenamefont{Schuller and Wohlfarth}(2004)}]{wohlfarth}
\bibinfo{author}{\bibfnamefont{F.~P.} \bibnamefont{Schuller}} \bibnamefont{and}
  \bibinfo{author}{\bibfnamefont{M.~N.~R.} \bibnamefont{Wohlfarth}},
  \bibinfo{journal}{Nucl. \ Phys. \ B} \textbf{\bibinfo{volume}{698}},
  \bibinfo{pages}{319} (\bibinfo{year}{2004}), \eprint{hep-th/0403056}.

\bibitem[{\citenamefont{Magnano and Sokolowski}(1994)}]{potentialsd}
\bibinfo{author}{\bibfnamefont{G.}~\bibnamefont{Magnano}} \bibnamefont{and}
  \bibinfo{author}{\bibfnamefont{L.~M.} \bibnamefont{Sokolowski}},
  \bibinfo{journal}{Phys. \ Rev. \ D} \textbf{\bibinfo{volume}{50}},
  \bibinfo{pages}{5039} (\bibinfo{year}{1994}), \eprint{gr-qc/9312008}.

\bibitem[{\citenamefont{Mignemi and Wiltshire}(1992)}]{mignemi}
\bibinfo{author}{\bibfnamefont{S.}~\bibnamefont{Mignemi}} \bibnamefont{and}
  \bibinfo{author}{\bibfnamefont{D.~L.} \bibnamefont{Wiltshire}},
  \bibinfo{journal}{Phys. \ Rev. \ D} \textbf{\bibinfo{volume}{46}},
  \bibinfo{pages}{1475} (\bibinfo{year}{1992}).

\bibitem[{\citenamefont{Misner et~al.}(1973)\citenamefont{Misner, Thorne, and
  Wheeler}}]{gravitation}
\bibinfo{author}{\bibfnamefont{C.~W.} \bibnamefont{Misner}},
  \bibinfo{author}{\bibfnamefont{K.~S.} \bibnamefont{Thorne}},
  \bibnamefont{and} \bibinfo{author}{\bibfnamefont{J.~A.}
  \bibnamefont{Wheeler}}, \emph{\bibinfo{title}{Gravitation}}
  (\bibinfo{publisher}{W. H. Freeman and Company}, \bibinfo{year}{1973}).

\bibitem[{\citenamefont{Weinberg}(1972)}]{grweinberg}
\bibinfo{author}{\bibfnamefont{S.}~\bibnamefont{Weinberg}},
  \emph{\bibinfo{title}{Gravitation and Cosmology}} (\bibinfo{publisher}{John
  Wiley and Sons}, \bibinfo{year}{1972}).

\bibitem[{\citenamefont{Kusmartsev
  et~al.}(1991{\natexlab{a}})\citenamefont{Kusmartsev, Mielke, and
  Schunck}}]{KMS91a}
\bibinfo{author}{\bibfnamefont{F.~V.} \bibnamefont{Kusmartsev}},
  \bibinfo{author}{\bibfnamefont{E.~W.} \bibnamefont{Mielke}},
  \bibnamefont{and} \bibinfo{author}{\bibfnamefont{F.~E.}
  \bibnamefont{Schunck}}, \bibinfo{journal}{Phys. \ Rev. \ D}
  \textbf{\bibinfo{volume}{43}}, \bibinfo{pages}{3895}
  (\bibinfo{year}{1991}{\natexlab{a}}).

\bibitem[{\citenamefont{Kusmartsev
  et~al.}(1991{\natexlab{b}})\citenamefont{Kusmartsev, Mielke, and
  Schunck}}]{KMS91b}
\bibinfo{author}{\bibfnamefont{F.~V.} \bibnamefont{Kusmartsev}},
  \bibinfo{author}{\bibfnamefont{E.~W.} \bibnamefont{Mielke}},
  \bibnamefont{and} \bibinfo{author}{\bibfnamefont{F.~E.}
  \bibnamefont{Schunck}}, \bibinfo{journal}{Phys. \ Lett. \ A}
  \textbf{\bibinfo{volume}{157}}, \bibinfo{pages}{465}
  (\bibinfo{year}{1991}{\natexlab{b}}).

\bibitem[{\citenamefont{Matos et~al.}(2000)\citenamefont{Matos, Guzm{\'a}n, and
  N{\'u}{\~n}ez}}]{matos2000a}
\bibinfo{author}{\bibfnamefont{T.}~\bibnamefont{Matos}},
  \bibinfo{author}{\bibfnamefont{F.~S.} \bibnamefont{Guzm{\'a}n}},
  \bibnamefont{and}
  \bibinfo{author}{\bibfnamefont{D.}~\bibnamefont{N{\'u}{\~n}ez}},
  \bibinfo{journal}{Phys. \ Rev. \ D} \textbf{\bibinfo{volume}{62}},
  \bibinfo{pages}{061301} (\bibinfo{year}{2000}), \eprint{astro-ph/0003398}.

\bibitem[{\citenamefont{Sokolowski}(1995)}]{sokol95}
\bibinfo{author}{\bibfnamefont{L.~M.} \bibnamefont{Sokolowski}},
  \emph{\bibinfo{title}{Universality of einstein's general relativity}}
  (\bibinfo{year}{1995}), \bibinfo{note}{plenary talk at the 14th Conference on
  General Relativity and Gravitation, Florence, 1995}, \eprint{gr-qc/9511073}.

\bibitem[{\citenamefont{Alcubierre et~al.}(2004)\citenamefont{Alcubierre,
  Gonz{\'a}lez, and Salgado}}]{ulises}
\bibinfo{author}{\bibfnamefont{M.}~\bibnamefont{Alcubierre}},
  \bibinfo{author}{\bibfnamefont{J.~A.} \bibnamefont{Gonz{\'a}lez}},
  \bibnamefont{and} \bibinfo{author}{\bibfnamefont{M.}~\bibnamefont{Salgado}},
  \bibinfo{journal}{Phys. \ Rev. \ D} \textbf{\bibinfo{volume}{70}},
  \bibinfo{pages}{064016} (\bibinfo{year}{2004}), \eprint{gr-qc/0403035}.

\bibitem[{\citenamefont{Guzm{\'a}n and Ure{\~n}a-L{\'o}pez}(2004)}]{laul2004a}
\bibinfo{author}{\bibfnamefont{F.~S.} \bibnamefont{Guzm{\'a}n}}
  \bibnamefont{and} \bibinfo{author}{\bibfnamefont{L.~A.}
  \bibnamefont{Ure{\~n}a-L{\'o}pez}}, \bibinfo{journal}{Phys. \ Rev. \ D}
  \textbf{\bibinfo{volume}{69}}, \bibinfo{pages}{124033}
  (\bibinfo{year}{2004}), \eprint{gr-qc/0404014}.

\bibitem[{\citenamefont{Belluci et~al.}(2001)\citenamefont{Belluci, Faraoni,
  and Babusci}}]{belluci}
\bibinfo{author}{\bibfnamefont{S.}~\bibnamefont{Belluci}},
  \bibinfo{author}{\bibfnamefont{V.}~\bibnamefont{Faraoni}}, \bibnamefont{and}
  \bibinfo{author}{\bibfnamefont{D.}~\bibnamefont{Babusci}},
  \bibinfo{journal}{Phys. \ Lett. \ A} \textbf{\bibinfo{volume}{282}},
  \bibinfo{pages}{357} (\bibinfo{year}{2001}), \eprint{hep-th/0103180}.

\bibitem[{\citenamefont{Maggiore and Nicolis}(2000)}]{maggiore}
\bibinfo{author}{\bibfnamefont{M.}~\bibnamefont{Maggiore}} \bibnamefont{and}
  \bibinfo{author}{\bibfnamefont{A.}~\bibnamefont{Nicolis}},
  \bibinfo{journal}{Phys. \ Rev. \ D} \textbf{\bibinfo{volume}{62}},
  \bibinfo{pages}{024004} (\bibinfo{year}{2000}), \eprint{gr-qc/9907055}.

\bibitem[{\citenamefont{Schunck and Mielke}(2003)}]{SM03}
\bibinfo{author}{\bibfnamefont{F.~E.} \bibnamefont{Schunck}} \bibnamefont{and}
  \bibinfo{author}{\bibfnamefont{E.~W.} \bibnamefont{Mielke}},
  \bibinfo{journal}{Class. \ Quantum \ Grav.} \textbf{\bibinfo{volume}{20}},
  \bibinfo{pages}{R301} (\bibinfo{year}{2003}).

\bibitem[{\citenamefont{Schunck and Torres}(2000)}]{ST00}
\bibinfo{author}{\bibfnamefont{F.~E.} \bibnamefont{Schunck}} \bibnamefont{and}
  \bibinfo{author}{\bibfnamefont{D.~F.} \bibnamefont{Torres}},
  \bibinfo{journal}{Int. J. Mod. Phys.} \textbf{\bibinfo{volume}{D9}},
  \bibinfo{pages}{601} (\bibinfo{year}{2000}), \eprint{gr-qc/9911038}.

\end{thebibliography}

\end{document}